\documentclass[twocolumn,showpacs,aps,prl,a4paper]{revtex4}
\usepackage{amsmath,graphicx}

\begin{document}

\title{Atoms and molecules in lattices:
condensates built on a shared vacuum}
\author{Tilman Esslinger$^1$ and Klaus M\o lmer$^2$}
\affiliation{1. Institute for Quantum Electronics, ETH  Zurich\\
CH-8093 Zurich, Switzerland\\
2. QUANTOP, Department of Physics and Astronomy,
University of Aarhus \\
DK-8000 \AA rhus C, Denmark}

\begin{abstract}
In optical lattices where each site is occupied in its lowest energy
state by a superposition of zero, one and two atoms, one can in a
controllable manner convert the atomic pair into a molecule while
retaining the vacuum and one-atom amplitudes.
The microscopic quantum coherence on each site
between the vacuum and the single molecule component leads
to a macroscopically populated molecular condensate when the lattice
is removed.
\end{abstract}


\maketitle

In this paper we present a simple quantum optical analysis of
the diffraction pattern of atoms and molecules released from a periodic
potential. We show that removal of atom
pairs from a lattice with low filling fraction may be used to
reestablish a robust atomic interference pattern after dephasing
has taken place due to collisional interaction. If the removal occurs
by photoassociation, a molecular condensate may be produced and
detected when the
lattice is removed. We do not make use of the Mott-insulator phase
transition dynamics \cite{Jaksch98,Jaksch02}.
On the contrary, the existence of a superposition
of vacuum and one or several atoms or molecules on every lattice
site is crucial, and our proposal works ideally in the absense of
atomic interactions.

Assume that an atomic Bose-Einstein condensate has been exposed to
a lattice potential, which was turned on slowly enough for the
spatial motion of the atoms to remain in the lowest energy state
but too fast for the Mott insulator transition to have happened.
This system is described as a single condensate in the $q=0$ state
of the lowest Bloch band of the potential, which is, in case of a
deep potential modulation, a superposition of Wannier states
localized in each well. We assume that all the atoms in the $m'th$
well are well described by the $m'th$ Wannier function
$\phi_m(x)=\phi_0(x-mL)$.

If the entire many body wavefunction is described as a coherent
state, i.e., an eigenstate of the atomic annihilation operators
$\hat{\Psi}(x)$, the state remains a coherent state, which factors
exactly into a product state of coherent states populating the
Wannier states, i.e., each site
is populated by a superposition $\sum_n c_n|n\rangle_m$ of zero, one, or
more atoms, where  the amplitudes $c_n$ can be
parametrized by a single coefficient $\beta$:
$c_n=\exp(-|\beta|^2/2)\beta^n/\sqrt{n!}$.
Such superposition states do not reflect the conservation of atom number.
The fluctuations in atom number on each lattice site, however, reflects
the entanglement of the
wells due to the splitting of the atoms - and it is a remarkable
feature of coherent states that this entanglement is well accounted for
by a product state (because a formal projection on a total number
eigenstate with very many atoms, which indeed produces a non-separable
state, does not change our physical predictions).

Since atoms are bound to occupy the ground states of the wells of the
lattice,
it is convenient to introduce the discrete set of atomic field
operators that removes atoms from the Wannier mode functions rather
than from a specific location in space. These are defined as
\begin{equation}
a_m = \int \phi_m(x) \hat{\Psi}(x) dx,
\end{equation}
and they obey the standard commutator relations
$[a_{m},a^{\dagger}_{m'}]= \delta_{m,m'}$.

With the product wave function $|\Psi\rangle = \Pi_m(\sum_n
c_n|n\rangle_m)$, the mean value of the mode annihilation
operator in each well equals
$\alpha := \langle a_m\rangle = \sum_{n=1}^{\infty}
c_{n-1}^*c_n\sqrt{n}$, the
number of atoms in each well equals $n := \langle a^{\dagger}_ma_m\rangle=
\sum_{n=1}^{\infty} |c_n|^2 n$,
and we have the inter-well coherence
$\langle a_{m'}^{\dagger}a_m\rangle = \alpha^*\alpha$ for $m \ne m'$.
The short range interaction
between atoms is a weak perturbation which causes a phase evolution
of the $c_n$ amplitudes with a frequency $\frac{U}{2}n(n-1)$. In the
absense of tunneling between wells, the product form of the wave
function remains exact, but the phase evolution leads to a reduction
of the atomic field amplitude $\alpha$  since the terms
$c_{n-1}^*c_n\sqrt{n}$ acquire different complex phases.
The resulting disappearance of atomic interference has been observed
in experiments \cite{Munichdiff,Yalediff}, and subsequent revivals
when the phase differences
reach multiples of $2\pi$  have also been shown \cite{Munichrevival}.

We now turn to the special case of on the order of or less than
unit mean occupancy of each well. The amplitudes on states with
three or more atoms are small and we shall mainly focus on the
lower $n$ components. Since $|c_0^*c_1+c_1^*c_2\sqrt{2}|$ may be
smaller than $|c_0^*c_1|$, it seems an interesting experiment to
remove the component with two atoms from the  state vector of each
well. This can be done by a carefully designed photoassociation
process \cite{Jaksch02,Javanainen,Heinzen00}, making use of the
fact that the two-atom component is in a fully determined initial
state. Neglecting higher atomic occupancies, the resulting state
vector becomes a product of two-species superposition states:
\begin{equation}
|\Psi\rangle = \Pi_m (c_0|{\mbox{vacuum}}\rangle + c_1|{\mbox{1
atom}}\rangle + c_2|{\mbox{1 molecule}}\rangle)_m.
\label{species}
\end{equation}

In the state (\ref{species}) we can readily determine the
mean atomic amplitude and population,
$\alpha=c_0^*c_1$ and $n=|c_1|^2$. The mean amplitude
acquires a finite robust value,
which is not corrupted by the atom-atom interaction before or after the
photoassociation. We may thus wait arbitrarily long before
we remove the atomic pairs, and still the coherence reappears.
Technically, we define condensate population as the largest eigenvalue
of the one-body density matrix, and applying the basis of Wannier
states, this matrix has the mean occupancy of the wells $n$ in the diagonal, and
the squared amplitude $|\alpha|^2$ in all other positions. In a system
with $N$ wells,
the eigenvector with equal amplitude on each Wannier function has the
largest eigenvalue of $(N-1)|\alpha|^2+n$, i.e., in the limit of
large $N$ the condensate fraction is $|\alpha|^2/n$. This
is also the value predicted by the
off-diagonal long-range order, $\langle a^{\dagger}_ma_{m'}\rangle
= |\alpha|^2 = n\cdot(|\alpha|^2/n)$. It is interesting
to note that in cases where $c_2$ is sufficiently dephased,
photoassociation increases the value of $|\alpha|^2$ and hence
{\it both the total occupancy of the condensate and the condensate fraction
increase by the removal of atoms}. The state before photoassociation is
a perfect, pure quantum state of the system, and it is related to
the perfect condensate state by a unitary process - but
the system cannot be described by a Gross-Pitaevskii wave function,
and the one-body density matrix cannot distinguish a deterministically
evolved phase from, e.g., a finite temperature decoherence effect.

Let us now turn our attention to the molecules prepared in the lattice.
It is clear from the state (\ref{species}), that the atomic and
molecular states are quite equivalent: there is a molecular population
$n_M = |c_2|^2$ per lattice site, and there is a mean molecular field
of $\alpha_M = c_0^*c_2$. The above analysis of condensate population
and fraction thus applies to the molecules if $n$ and $\alpha$ are
replaced by $n_M$ and $\alpha_M$. The preparation of molecular
condensates has received quite some interest, in particular with the
recent experiments at JILA \cite{Claussen}, where coherent oscillations
between an atomic condensate and a condensate of very loosely
bound molecules are observed and agree well with a detailed theory
of collision dynamics near a Feschback resonance \cite{Holland},
see also \cite{Burnettmol}.
In addition to the detailed binary collision dynamics, it is
necessary to represent correctly the many-body dynamics in the case
where many atoms are converted into molecules. In comparison, our
analysis is much simpler since there is either nothing or just a single
atom or just a single molecule in every lattice site (as long as the
truncation of the atomic state above two atoms is a valid
approximation). There is accordingly just a single population
independent coupling between the relevant states
$|n=2\ \ {\mbox{atoms}}\rangle$
and $|n_M=1\ \ {\mbox{molecule}}\rangle$.
From an experimental point of view our molecules
can be prepared in a deeper bound molecular state,
selected coherently by the coupling laser fields
rather than by Feshbach collision dynamics, and they are
prepared in a spatial condensate for which the phase coherence
can be readily detected
by diffraction experiments similar to the ones  applied to atoms
in \cite{Munichdiff}, see below.

Our proposal may also be compared with the suggestion by Jaksch et
al. \cite{Jaksch02}, where the Mott-insulator dynamics is used to
prepare a lattice with precisely two atoms per site, these atoms
are transferred into a molecule, and the molecular Mott-insulator
is subsequently melted to yield the molecular condensate. Our
proposal emphasizes a simple but spectacular phenomenon: a
zero-quantum amplitude is enough to yield a mean field, and hence
the condensate establishes itself simply by the production of
molecules out of the two-atom  component of the atomic state. This
is so, because the atomic vacuum is of course also a molecular
vacuum (as well as a vacuum for larger molecules, butterflies and
freight trains). On the experimental side, the molecular
condensate may ideally be created from an atomic condensate of
non-interacting particles, and we only need to transfer the atoms
slowly enough into the lattice to have a well-specified initial
state for the photo-association process.

The analysis is quite simple and fully analytical in the limit of very
weakly populated sites. Also for a larger mean number of atoms on each
site the conversion into
molecules will lead to a molecular mean field, i.e., a molecular
condensate. To investigate this, we have solved numerically the
time evolution of the simple photoassociation Hamiltonian
\begin{equation}
H_{PA} = \sum_m\chi(a_m^2b_m^{\dagger}+(a_m^{\dagger})^2b_m)
\label{HPA}
\end{equation}
where $b_m$ is the operator of annihilation of a molecule at site
$m$. Starting with a coherent atomic state, the Hamiltonian
(\ref{HPA}) introduces in every lattice well a superposition state
$\sum c_{n,n_M}(t)|n,n_M\rangle$, from which the mean number of
molecules and the mean molecular field is readily calculated. The
results of such calculation are shown in Fig.1. It shows the
number of molecules and the absolute square of the molecular field
as functions of time: the conversion is oscillatory in time, and
to optimize the coherent molecular component, one should stop the
photoassociation process after a time interval depending on the
number of atoms per lattice site. For simplicity we did not
incorporate atom-atom interactions or interactions between the
atoms and molecules in these calculations since our main focus is
on lattices with low population.

\begin{figure}[b]
  \centering
 \includegraphics[width=7cm]{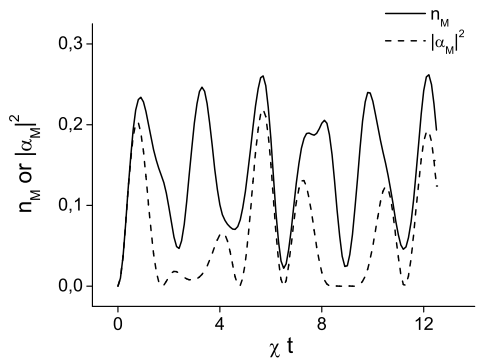}
\caption{Incoherent and coherent molecular components prepared by
photoassociation of atoms. The number of molecules (solid line)
and the squared norm of the mean molecular field (dashed line) as
a function of time is shown, starting at time zero with a coherent
atomic state with unit mean number.}
  \label{figure}
\end{figure}

In the above analysis we have shown that the on-site mean field
amplitude and population mathematically determine the condensate fraction.
These quantities are directly measurable by the diffraction occuring
when the particles are released from the lattice. The detection of
the diffraction pattern has precisely been the hall mark of coherence
\cite{Munichdiff,Yalediff}, and coherence decay and revivals
\cite{Munichrevival} in experiments, and we
present here a brief theoretical description, linking the diffraction pattern
directly to our quantities $n$ and $\alpha$.

If one introduces a complete and orthonormal set of wavefunctions
$\{\phi_{m,l}\}$, where $\phi_{m,0}=\phi_m$ are the Wannier functions
applied above, and $\{\phi_{m,l}\}$ with $l>0$ are higher excited states
centered on the $m^{th}$ lattice site, the commutator relations
$[\hat{\Psi}(x),\hat{\Psi}^{\dagger}(x')]=\delta(x-x')$,
lead to  $[a_{m,l},a^{\dagger}_{m',l'}]=
\delta_{m,m'}\delta_{l,l'}$, with
the mode annihilation operators defined
through
\begin{equation}
a_{m,l}=\int \phi_{m,l}(x) \hat{\Psi}(x) dx.
\label{annihil}
\end{equation}
We shall need expressions for  the position dependent
annihilation operators in terms of the mode functions,
\begin{eqnarray}
\hat{\Psi}(x)=\sum_{m,l} a_{m,l} \phi_{m,l}(x)\nonumber \\
=a_{\overline{m}} \phi_{\overline{m}}(x) + \delta\hat{\Psi}(x),
\label{one-mode}
\end{eqnarray}
where in the second line we include only explicitly the Wannier state
$\phi_{\overline{m}}$, where $\overline{m}$
enumerates the well which is most close to the position coordinate $x$,
and where the 'noise-term' $ \delta\hat{\Psi}(x)$ represents all
other modes, which are
either unoccupied $(l > 0)$ or which have no support at the location
$x$, ($m \ne \overline{m}$).
This noise term does not contribute to atom counting signals.

In experiments on trapped condensates, the spatial properties  of
the atomic cloud are detected by releasing the cloud and
letting it expand. In case of the weakly filled lattices, one may
neglect interactions during the expansion, and the final spatial
distribution hence maps the momentum distribution of the atoms in the
lattice.
We determine the momentum distribution by writing the atomic
field operator that annihilates an atom with momentum $\hbar k$,
in terms of the spatial field operators:
\begin{eqnarray}
\hat{\Psi}(k) = \frac{1}{\sqrt{2\pi}}\int dx e^{ikx} \hat{\Psi}(x)\nonumber
\\ = \sum_{\overline{m}}a_{\overline{m}}e^{ik\overline{m}L}\phi(k) + \delta
\hat{\Psi}(k),
\label{annihilk}
\end{eqnarray}
where $\phi(k)$ is the Fourier transfom of the Wannier function $\phi_0(x)$,
and where $\delta \hat{\Psi}(k)$ is a vacuum noise field with zero mean and
zero intensity.  Although the quantity ${\overline{m}}$ originally
appeared as a complicated index, depending on $x$, the integral
over all positions leads to a sum over all positive and negative
integer values for ${\overline{m}}$, i.e., it has become a conventional
summation index. The number of atoms with momentum $\hbar k$ is
given by the expectation value of the operator $\hat{\Psi}^{\dagger}(k)
\hat{\Psi}(k)$:
\begin{eqnarray}
\langle \hat{\Psi}^{\dagger}(k) \hat{\Psi}(k)\rangle =
|\phi(k)|^2\sum_{m,m'}e^{i(m-m')kL}\langle a_{m'}^{\dagger}a_m\rangle.
\end{eqnarray}
Thus
\begin{eqnarray}
\label{pattern}
\langle \hat{\Psi}^{\dagger}(k) \hat{\Psi}(k)\rangle \nonumber \\
 = N|\phi(k)|^2\left( (n-|\alpha|^2)+
|\alpha|^2\sum_q \delta(k-q\frac{2\pi}{L})\right),
\label{distribution}
\end{eqnarray}
where $N$ is the (large) number of lattice sites. The sum over $q$
gives rise to a comb at lattice momenta $q\frac{2\pi}{L}$
with a modulation proportional to the square of the mean field
amplitude. This comb sits on top of a flat background caused by
the incoherent population of the wells, and the whole distribution
is comprised within the width of the single well momentum distribution
$|\phi(k)|^2$.

A diffraction pattern in the molecular distribution is the clear
signature of a molecular condensate. By counting the released
atoms or the molecules on a position sensitive detector the values
of $n$ and $\alpha$ and thus the condensate fraction can by
determined.

In summary, we have presented a method which by photoassociative
removal of atoms can at the same time
return an atomic system to a state with a macroscopic population of
a single quantum state and prepare a molecular
system with a similar macroscopic population.
Weak fields owe their mean amplitude to the existence of
both vacuum and single quantum excitations, and the molecular condensate
exists due to the molecular vacuum component being already populated in
the atomic condensate.

Eq.(\ref{pattern}) relates the diffraction pattern to the mean field
amplitudes existing prior to the detection.
Two independently prepared condensates
also show interference \cite{Ketterlediff}.
It has been demonstrated theoretically, \cite{Theorydiff}, that
this interference builds up as a consequence of the first random
detection events, and that one can think of the state of the two
condensates without initial mean fields as a statistical mixture
of states with different relative phases \cite{Klausdiff},
which all yield interference patterns but with different offsets.
This same mechanism is not at work when atoms or molecules emerge from a
large number of independent sites, since the uniform distribution of the phase
variable at every site has vanishing probability to produce
the regular progression of the phase over several sites needed
for a high visibility diffraction pattern.

Our approach is
quantum optical in nature: by carefully selected unitary operations
we generate from a given initial state a final desired state of the
system. This implies that the formal properties of the state vector of
the system are made to the design, but the dynamical properties of the
system may have only little in common with the properties usually
ascribed to a Bose-Einstein condensate. Since the state is not the
ground state or the
thermodynamic equilibrium state under an applied Hamiltonian, it will
not act in the same way as such a state under external perturbations,
cf. for example the issue of superfluidity, discussed recently in a similar
physical set-up as ours \cite{Roth}. Our approach, however, suggests the
possibility to create an approximation to the ground state or
thermal equilibrium state of a given Hamiltonian, which can be subsequently
turned on and in this way prepare states that might otherwise
only be produced after prohibitively long time scales.

We would like to acknowledge the hospitality of the European
Centre for Theoretical Studies in Nuclear Physics and Related
Areas (ECT) during the Summer Program on Bose-Einstein
condensation.

\end{document}